\documentclass[usenatbib]{mn2e}

\usepackage{epsfig}
\usepackage{amsmath}
\newcommand{\beq}{\begin{equation}}
\newcommand{\eeq}{\end{equation}}

\def\ra#1h#2m#3.#4s{$\rm {RA} \ #1^h \ #2^m \ #3^s#4 \ $}
\def\dec#1d#2m#3s{$\rm {Dec.}\ +#1^{\circ} \ #2' \ #3''$}

\def\d#1by#2{{{\rm d}{#1} \over {\rm d}{#2}}}

\def\sigV{\langle \sigma V\rangle_{A}}

\def\sd{{\mathrm d}}

\def\spose#1{\hbox to 0pt{#1\hss}}
\def\lta{\mathrel{\spose{\lower 3pt\hbox{$\sim$}}
    \raise 2.0pt\hbox{$<$}}}
\def\gta{\mathrel{\spose{\lower 3pt\hbox{$\sim$}}
    \raise 2.0pt\hbox{$>$}}}

\title[Comptonisation of CMB Photons in Dwarf Spheroidal Galaxies]
{Comptonisation of CMB Photons in Dwarf Spheroidal Galaxies}

\author[Culverhouse et al.]{Thomas L. Culverhouse$^{1}$\thanks
{E-mail: tlc26@mrao.cam.ac.uk, nwe@ast.cam.ac.uk, cola@mporzio.astro.it},
N. Wyn Evans$^{2}$\footnotemark[1] \& S. Colafrancesco$^3$\footnotemark[1]\\
$^{1}$Astrophysics Group, Cavendish Laboratory,
      Madingley Road, Cambridge CB3 0HE\\
$^{2}$Institute of Astronomy, University of Cambridge,
      Madingley Road, Cambridge CB3 0HA\\
$^{3}$INAF --Osservatorio Astronomico di Roma, via Frascati 33, 0040
      Monteporzio, Italy\\
}

\begin{document}

\maketitle

\begin{abstract} We present theoretical modelling of the electron 
distribution produced by annihilating neutralino dark matter in dwarf
spheroidal galaxies (dSphs). In particular, we follow up the idea
of Colafrancesco (2004) and find that such electrons distort the
cosmic microwave background (CMB) by the Sunyaev-Zeldovich effect.
For an assumed neutralino mass of 10 GeV and beam size of $1''$, the
SZ temperature decrement is of the order of nano-Kelvin for dSph
models with a soft core. By contrast, it is of the order of
micro-Kelvin for the strongly cusped dSph models favoured by some
cosmological simulations. Although this is out of reach of current
instruments, it may well be detectable by future mm telescopes, such
as ALMA. We also show that the upscattered CMB photons have energies
within reach of upcoming X-ray observatories, but that the flux of
such photons is too small to be detectable soon. Nonetheless, we
conclude that searching for the dark matter induced Sunyaev-Zeldovich
effect is a promising way of constraining the dark distribution in
dSphs, especially if the particles are light.
\end{abstract}


\section{Introduction}
\label{sec:intro}
Dwarf spheroidal galaxies (dSphs) are important probes of dark
matter. They are among the highest mass-to-light systems known, and
the dynamics of their sparse stellar populations is governed by the
dominant dark matter distribution. In addition, no emission has been
detected from dSphs in wavebands other than the optical, indicating a
lack of internal dust or gas \citep{fg,bonanos}.

Here, we consider the distortion of the cosmic microwave background
(CMB) by the non-thermal population of secondary electrons generated
by dark matter annihilation. This is an example of the
Sunyaev-Zeldovich (SZ) effect~\citep{sz}. Since we are dealing with
electrons produced from dark matter annihilation, we write the
distortion as the dSZ effect. \cite{ek} and \cite{sc2003} calculated
the signal expected from the SZ effect of a relativistic plasma, while
\cite{sc2004} determined the dSZ effect in galaxy clusters.

As first pointed out in \cite{sc2004}, dSphs are attractive targets
because they have very high mass-to-light ratios and because they have
few contaminants. In particular, they have little or no internal
magnetic field, so it is not possible for synchrotron emission (from
annhilation electrons or otherwise) to contaminate the dSZ
effect. Since dSphs are believed to be almost devoid of interstellar
gas, other mechanisms such as HI or CO line emission will not be
important. CMB distortions are therefore a rather clean method to
detect the annihilation signature. We therefore use existing models
of the dark matter distribution in dSphs, and a possible form of the
energy spectrum of electrons produced by dark matter annihilation, to derive
the temperature change in the CMB. We focus on dark matter models explicitly
constrained by observations, rather than simulations.

For our predictions, we assume that the cold dark matter particle is
the lightest supersymmetric particle, the
neutralino~\citep{Jung}. Current limits on the neutralino mass
$M_{\chi}$ and centre-of-mass velocity-averaged cross-section $\sigV$
have been reviewed recently by \cite{bhs}. Based on these results, and
considerations of current or upcoming experiments, we investigate the
SUSY parameters $\sigV=10^{-26}\ \mathrm{cm^{3}s^{-1}}$ and
$M_{\chi}=10\ \mathrm{GeV}$. Such low mass particles may provide
a sizeable contribution to the matter density in the Universe
\citep{bottino}, and hence are worthy of consideration. However, some
dark matter candidates -- such as the neutralino in the most commonly
studied minimal supersymmetric models or the lightest Kaluza-Klein
particle -- must be more massive than this. The value assumed for
$\sigV$ is consistent with the expected relic density in a Universe
with $\Omega_{m}=0.3$ and $H_{0}=70\mathrm{km s^{-1} Mpc^{-1}}$
\citep{cm}.We derive the dependency of dSZ signal on $\sigV$ and
$M_{\chi}$, and show that the brightness temperature decrement
$\Delta T_{B}\propto\sigV M_{\chi}^{-3}$. We initially calculate the expected
signal for this optimistic choice of dark matter parameters.

The format of the paper is as follows. In Section \ref{sec:dsph}, we
discuss our dSph dark halo models, based on current work in the
literature. We then discuss possible products of dark matter particle
annihilation in Section \ref{sec:DManni}. Observational consequences
of such annihilation events are presented in Section
\ref{sec:simobs}. We conclude in Section \ref{sec:conclusions}.

\section{Dwarf Spheroidal Models}

\label{sec:dsph}

With spherical symmetry of the dark halo assumed, we use the results
of \citet{efs}, who fit observational data on the Draco dSph
(currently orbiting the Milky Way) from \citet{mark} to two sets of
models via the Jeans equation \citep{bt}.

\subsection{Cusped Halo Models}
\label{subsec:chm}

Cusped halo models (CHMs)are favoured by numerical simulations, as reported
by \citet{nfw} (NFW) and \citet{ben}). Using arguments based on the
survivability of kinematically cold substructure, \citet{jan} argued
against a cusped halo for at least the Ursa Minor dSph.  For the rest
of the dSphs, cusped halos remain viable.  We therefore consider
number density radial profiles of the form
\begin{equation}
n\left(\hat{r}\right)=n_{0}a\left(\hat{r}\right),
\label{eq:chm}
\end{equation}
where $n_{0}=Ar_{s}^{-3}/M_{\chi}$, and
$a\left(\hat{r}\right)=\hat{r}^{-\gamma}\left(1+\hat{r}\right)^{\gamma-3}$, 
with $\hat{r}=r/r_{s}$.  Table~\ref{tab:chmtable1} gives the cusp slope 
$\gamma$, the scale radius $r_{s}$, the tidal radius $r_{t}$, and the overall
normalisation $A$.  The model is truncated at $r_{t}$, whose value depends on
 the Milky Way dark halo model.
\begin{table} 
\begin{center}
\caption{List of CHM model parameters for the Draco dSph. Values for
$r_{t}$ in parentheses are for NFW models of the Milky Way, as opposed
to isothermal power-law models which are without brackets.
\label{tab:chmtable1}}
\begin{tabular}{ccccc}
\hline \hline
$\gamma$ & $\mathrm{A}\times 10^{7} M_{\odot}$ & $r_{s}/\mathrm{kpc}$ & $r_{t}/\mathrm{kpc}$ & $M(r_{t})/10^{8}M_{\odot}$ \\
\hline
0.5 & 2.3 & 0.32 & 6.6 (1.5) & 5.5\\
1.0 & 3.3 & 0.62 & 7.0 (1.6) & 6.6\\
1.5 & 2.9 & 1.0 & 6.5 (1.5) & 5.5\\
\hline \hline
\end{tabular}
\end{center}
\end{table}

To address the problem of the divergent central density in this model,
we use the arguments of \citet{bc} and \citet{tyler}. The density
profile is truncated at a radius $r_{\rm min}$, where we assume the
dark matter annihilation rate matches the collapse timescale of the
cusp. With this assumption, a small constant density core is created
with radius $r_{\rm min}$
\begin{equation}
r_{\rm min}=r_{t}\langle \sigma
V\rangle^{1/2}_{A}\left(\frac{n_{\rm min}}{4\pi G M_{\chi}}\right)^{1/4},
\label{eq:rmin1}
\end{equation}
where $n_{\rm min}$ is the number density at the location of the
constant density region. At small radii, Eq.(\ref{eq:chm}) reduces to
$n(r)\simeq AM^{-1}_{\chi}r^{-3}_{s}\hat{r}^{-\gamma}$, and hence
we find:
\begin{equation}
r^{1+\gamma/4}_{\rm min}=r_{t}\langle \sigma
V\rangle^{1/2}_{A}\left(\frac{Ar^{\gamma-3}_{s}}{4\pi G
M^{2}_{\chi}}\right)^{1/4}.
\label{eq:rmin2}
\end{equation}
Typically, $r_{\rm min}\sim 10^{13}\mathrm{m}$, and for such a small
value, it is unlikely that tidal forces (from M31 or the Milky Way)
could disrupt the central cusp. It should be noted that the Moore
profile ($\gamma=1.5$) represents an extreme case for the inner slope
of the cusp. Recent numerical simulations
\citep[e.g. ][]{nav04,diem04} point towards a milder density slope of
$\gamma\simeq1.1$. However, we retain the Moore profile here as the
models satisfy the observational constraints from stellar radial
velocities. In addition, this profile allows an upper limit of the
magnitude of the dSZ effect, enabling a broad yet well motivated
parameter space to be investigated.

\subsection{Cored Power-Law Models}
\label{subsec:cpl}

The second family of dark halo profiles studied are the cored
power-law (CPL) models \citep{evans}. Again, these satisfy the Draco
velocity dispersion observations, and take the form
\begin{equation}
n\left(\hat{r}\right)=n_{0}b\left(\hat{r}\right),
\label{eq:cpl}
\end{equation}
where $n_{0}=v^{2}_{a}/4\pi Gr^{2}_{c}$. The radial dependence is
given by
\begin{equation}
b\left(\hat{r}\right)=\frac{3+\hat{r}^{2}\left(1-\alpha\right)}
{\left(1+\hat{r}^{2}\right)^{2+\alpha/2}},
\label{eq:cpl2}
\end{equation}
where $\hat{r}=r/r_{c}$.  Table \ref{tab:cpltable1} gives the slope
$\alpha$, the core radius $r_{c}$, the tidal radius $r_{t}$ and the
velocity scale $v_{a}$. The dSph dark halo extends out to $r_{t}$, the
value of which again depends on the model (isothermal power-law or
NFW) used for the Milky Way.
\begin{table}
\begin{center}
\caption{List of CPL model parameters for the Draco dSph. Values for
$r_{t}$ in parentheses are for NFW models of the Milky Way, as opposed
to isothermal power-law models which are without brackets.
\label{tab:cpltable1}}
\begin{tabular}{ccccc}
\hline \hline
$\alpha$ & $v_{a}/ \mathrm{km\ s}^{-1}$ & $r_{c}/\mathrm{kpc}$ & $r_{t}/\mathrm{kpc}$ & $M(r_{t})/10^{8}M_{\odot}$ \\
\hline
0.2 & 24.7 & 0.25 & 6.2 (1.3) & 4.6\\
0 & 22.9 & 0.23 & 7.8 (1.4) & 9.5\\
-0.2 & 20.9 & 0.21 & 10.1 (1.6) & 22.43\\
\hline \hline
\end{tabular}
\end{center}
\end{table}
\section{Dark Matter Annihilation}
\label{sec:DManni}
Having established a set of representative dark matter halo models for
dSphs, which embrace a range of possible structures, we determine 
the decay products of annihilating neutralinos. Electrons produced in this
fashion have an associated cooling function $b\left(E\right)$,
comprised of synchrotron losses, inverse Comption scattering (ICS) and
Coulomb losses \citep{sc2004}:
\begin{equation}
b\left(E\right)=\left(\frac{\sd E}{\sd t}\right)_{\rm
  syn}+\left(\frac{\sd E}{\sd t}\right)_{\rm ICS}+\left(\frac{\sd E}{\sd
  t}\right)_{\rm Coul}\mathrm{\ GeV\ s^{-1}}.
\label{eq:bE}
\end{equation}
At electron energies $E\ge 150 \mathrm{MeV}$~\citep{bc}, this function
is dominated by the first two terms, so the Coulomb term may safely be
dropped from the cooling function. In addition, there is no evidence
in favour of significant magnetic fields in dSphs~\citep{fg}. We
therefore also drop the synchrotron cooling term, leaving~\citep{cm}
\begin{equation}
b\left(E\right)=2.5\times10^{-17}\times E^{2}\mathrm{\ GeV\ s^{-1}}.
\label{eq:bEhigh}
\end{equation}
Energy losses are efficient in the dSph halo, and dark matter
annihilation from infalling material continuously refills the electron
spectrum. In a first approximation, it seems reasonable to neglect
diffusive effects, as dark matter annihilation replenishes the
electron spectrum efficiently, at least in the central parts of the
dSph.

Furthermore, we consider the dark matter to maintain a constant
radial profile on the timescale of dark matter annihilation. This
means that any time-variation of the electron distribution is
negligible. The source function of electrons, $Q_{e}\left(E,r\right)$,
generated in this fashion therefore obeys the stationary
diffusion-loss equation (see, for example, Longair 1994)
\begin{equation}
\frac{\partial}{\partial E}\left(
\frac{\mathrm{d}n_{e}\left(E,r\right)}{\mathrm{d}E_{e}}
b\left(E\right)\right)=-Q_{e}\left(E,r\right),
\label{eq:diffloss}
\end{equation}
which we integrate to obtain the annihilation-produced electron energy
distribution $\mathrm{d}n_{e}\left(E,r\right)/\mathrm{d}E_{e}$:
\begin{equation}
\frac{\mathrm{d}n_{e}}{\mathrm{d}E_{e}}\left(E,r\right)
=-\frac{1}{b\left(E\right)}\int Q_{e}
\left(E,r\right) \mathrm{d}E.
\label{eq:intdiffloss}
\end{equation}

The source function itself results from annihilation products of
neutralino collisions. In this calculation, we follow the work of
\cite{tyler}, who used the Hill (1983) formula to determine the
densities of particles produced by dark matter annihilation. Quark
pairs and their subsequent fragmentation lead to pions as the main
annihilation products. The primary decay particles are neutral
pions, which decay to gamma rays, and charged pions, which decay as
\beq
\pi^{+}\rightarrow\mu^{+}\nu_{\mu}\ \mathrm{and}\
\pi^{-}\rightarrow\mu^{-}\bar{\nu}_{\mu}.
\label{eq:piondecay}
\eeq 
The muons then decay to electrons via \beq \mu^{+}\rightarrow
e^{+}\bar{\nu}_{\mu}\nu_{e}\ \mathrm{and}\ \mu^{-}\rightarrow
e^{-}\nu_{\mu}\bar{\nu}_{e}.
\label{eq:muondecay}
\eeq 
The number spectrum of electrons from a single $\chi\chi$
annihilation is then given by:
\beq
\frac{\mathrm{d}N_{e}}{\mathrm{d}E_{e}}=\int^{M_{\chi}}_{E_{e}}
\int^{E_{\mu}/\bar{r}}_{E_{\mu}}W_{\pi}\frac{\mathrm{d}N^{(\pi)}_{\mu}}
{\mathrm{d}E_{\mu}}\frac{\mathrm{d}N^{(\mu)}_{e}}{\mathrm{d}E_{e}}
\mathrm{d}E_{\pi}\mathrm{d}E_{\mu},
\label{eq:dnde}
\eeq
where $\bar{r}\equiv \left(m_{\mu}/m_{\pi}\right)^{2}$, and the charged 
pion multiplicity per annihilation event is
\beq
W_{\pi}=\frac{4}{3}\frac{15}{16M_{\chi}}\left(\frac{E_{\pi}}{M_{\chi}}\right)^{-3/2}
\left(1-\frac{E_{\pi}}{M_{\chi}}\right)^{2},
\label{eq:wpi}
\eeq
where the factor of $4/3$ accounts for the fact that annihilation 
electrons are only produced by charged pions, and that quarks (which eventually
decay to charged pions) are produced in pairs. The number spectrum of muons 
produced per charged pion decay is
\beq
\frac{\mathrm{d}N^{\pi}_{\mu}}{\mathrm{d}E_{\mu}}=\frac{1}{E_{\pi}}
\frac{m^{2}_{\pi}}{m^{2}_{\pi}-m^{2}_{\mu}},
\label{eq:dnmudemu}
\eeq
and
\beq
\frac{\mathrm{d}N^{\mu}_{e}}{\mathrm{d}E_{e}}=\frac{2}{E_{\mu}}
\left(\frac{5}{6}-\frac{3}{2}\left(\frac{E_{e}}{E_{\mu}}\right)^{2}+
\frac{2}{3}\left(\frac{E_{e}}{E_{\mu}}\right)^{3} \right)
\label{eq:dnedee}
\eeq
is the number spectrum of electrons per muon decay.
After some algebra, with the above forms for the decay product energy spectrum,
 Eq.(\ref{eq:dnde}) has an analytic solution:
\begin{multline}
\frac{\mathrm{d}N_{e}}{\mathrm{d}E_{e}}=\frac{15}{8M_{\chi}}\frac{m^{2}_{\pi}}
{m^{2}_{\pi}-m^{2}_{\mu}}\times \Biggl[
c_{1}z^{-3/2}+ \\c_{2}z^{-1/2}+ c_{3}+c_{4}z^{1/2}+c_{5}z^{2}
+c_{6}z^{3}\Biggr],
\label{eq:analdnde}
\end{multline}
in units of $\mathrm{GeV}^{-1}$, where $z=E_{e}/M_{\chi}$.  The coefficients 
$c_{i}$ are $\left(0.1039,-1.2218,2.4800,-1.5406,0.2205,-0.04197\right)$. 
Scaling this expression, we arrive at the source function 
\begin{multline}
Q_{e}\left(E,r\right)=\frac{4}{3}\frac{15}{8}\frac{m^{2}_{\pi}}
{m^{2}_{\pi}-m^{2}_{\mu}}\left(\frac{M_{\chi}}
{\mathrm{GeV}}\right)^{-1}\left(\frac{n_{\chi}}{\mathrm{cm}^{-3}}\right)^{2}\\\left(\frac{\sigV}{10^{-26}\mathrm{cm^{3}\ s^{-1}}}\right)
\times \Biggl[c_{1}z^{-3/2}+c_{2}z^{-1/2}\\
+ c_{3}+c_{4}z^{1/2}+c_{5}z^{2}
+c_{6}z^{3}\Biggr],
\label{eq:tylersource}
\end{multline}
where the appropriate units are $\mathrm{GeV}^{-1}\mathrm{cm}^{-3}
\mathrm{s}^{-1}$.
This needs to be multiplied by a further factor of 2 to account for
the contributions of electrons and positrons. Using this expression,
and Eq.(\ref{eq:intdiffloss}), we arrive at the equilibrium electron
energy distribution
\begin{multline}
\frac{\mathrm{d}n_{e}}{\mathrm{d}E_{e}}=4.6848\times10^{-9}
\left(\frac{M_{\chi}}
{\mathrm{GeV}}\right)^{-2}\left(\frac{n_{\chi}}{\mathrm{cm}^{-3}}\right)^{2}\\
\left(\frac{\sigV}{10^{-26}\mathrm{cm^{3}\ s^{-1}}}\right)
\times\Biggl[d_{1}z^{-5/2}+d_{2}z^{-3/2}+\\ d_{3}z^{-1} +d_{4}z^{-1/2}
+d_{5}z +d_{6}z^{2}\Biggr],
\label{eq:tylerne}
\end{multline}
in units of $\mathrm{GeV}^{-1}\mathrm{cm}^{-3}$. 
\begin{figure}
\begin{center}
\includegraphics[width=7.0cm,angle=270]{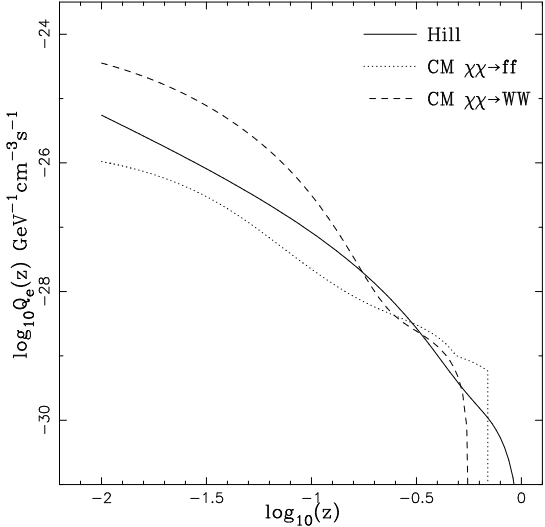} 
\caption[dummy]{Electron source function, as described by
Eq.\ref{eq:tylersource} (denoted Hill), assuming $M_{\chi}=100\ \mathrm{GeV}$,
$\sigV=10^{-26}\ \mathrm{cm^{3}s^{-1}}$ and
$n_{\chi}=1\mathrm{cm}^{-3}$. Also shown are the source functions described in 
Appendix A of \cite{cm}
, i.e. electrons produced in fermion-dominated 
annihilation $\chi\chi\rightarrow ff$, and gauge-boson dominated annihilation,
$\chi\chi\rightarrow WW$.}
\label{fig:tylersource}
\end{center}
\end{figure}
\begin{figure}
\begin{center}
\includegraphics[width=7.0cm,angle=270]{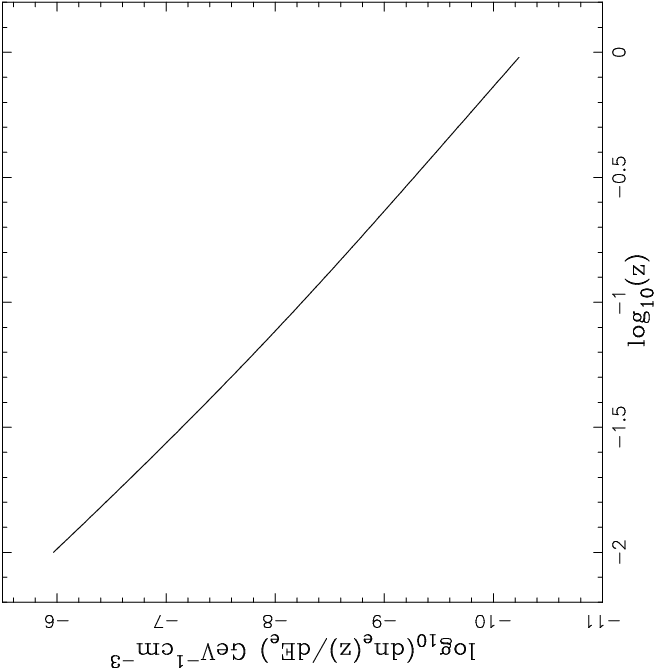}
\caption{Equilibrium electron energy distribution, calculated from
Eq.\ref{eq:tylerne}, using $M_{\chi}=10\ \mathrm{GeV}$, $\sigV=10^{-26}\
\mathrm{cm^{3}s^{-1}}$ and $n_{\chi}=1\mathrm{cm}^{-3}$.}
\label{fig:dnde}
\end{center}
\end{figure}
In this expression, the radial dependence of
$\mathrm{d}n_{e}/\mathrm{d}E_{e}$ arises in the $n_{\chi}$
factors. The new coefficients
$d_{i}=\left(0.2078,2.4436,-2.4800,1.0271,-0.07333,0.01049\right)$. 
Equations (\ref{eq:tylersource}) and (\ref{eq:tylerne}) are displayed 
in Figs \ref{fig:tylersource} and \ref{fig:dnde} respectively.

For completeness, in Fig~\ref{fig:tylersource} we also display
two other possible source functions presented in \cite{cm}. These
source functions arise from fermion-dominated annihilation in the
first instance (denoted CM $\chi\chi\rightarrow ff$), and gauge boson
dominated annihilation (CM $\chi\chi\rightarrow WW$) in the second. At
the crucial low energy end of the spectrum, the three separate source
functions exhibit amplitudes within 1.5 orders of magnitude, and
similar power-law slopes. This implies that our calculations may
increase or decrease by approximately one order of magnitude in either
direction, depending on the precise details of the source function. As
the results of \cite{cm} were derived in an independent manner to
those of \cite{hill}, our results may also hold for neutralino
compositions other than those considered here. The crucial quantity
for the magnitude of the dSZ effect is the lower limit of electron
energy, which can increase rapidly for any power-law source function
$Q_{e}\sim z^{-\beta}$, where $\beta>0$ is a generic slope
parameter. It is highly unlikely on energetic grounds that electrons
produced by annihilating dark matter can have $Q_{e}$ rising with
$z$. Although more complicated forms for $Q_{e}$ from general electron
sources are possible, such as a double power-law \citep{cm}, these
again have negative slopes and therefore it is the lowest electron
energies that are most significant.

From Eq (\ref{eq:tylerne}), we may calculate the number density of 
electrons in the halo:
\begin{equation}
n_{e}=\int^{M_{\chi}}_{0.01M_{\chi}} 
\frac{\mathrm{d}n_{e}}{\mathrm{d}E_{e}}\mathrm{d}E_{e}.
\label{eq:ndensity}
\end{equation}
The limits are chosen to be $0.01M_{\chi}<E_{e}<M_{\chi}$, based on
the analysis of \cite{KamionkowskiT}, who calculated the positron
source function in a similar manner to that described above. We will
see that the upper limit is somewhat irrelevant, as the source function
falls rapidly to zero as the electron energy approaches the neutralino
rest mass energy.

Evaluating this leads to
\begin{equation}
n_{e}=8.09\times10^{-7}\left(\frac{n_{\chi}}{\mathrm{cm}^{-3}}\right)^{2}
\left(\frac{\sigV}{10^{-26}\mathrm{cm^{3}\ s^{-1}}}\right)\left(\frac{M_{\chi}}
{\mathrm{GeV}}\right)^{-1},
\label{eq:ner}
\end{equation}
in units of $\mathrm{\ cm^{-3}}$.

We now represent Eq~(\ref{eq:tylerne}) in terms of its momentum
spectrum, as this quantity will be used later to calculate the
frequency spectrum of upscattered CMB photons. Specifically, we
write
\begin{equation}
\frac{\mathrm{d}n_{e}}{\mathrm{d}p}=n_{e}\left(r\right)f_{e}\left(p\right),
\label{eq:npspec}
\end{equation}
where $p=\beta_{e}\gamma_{e}$ is the normalised electron momentum, and 
$f_{e}\left(p\right)$ has the property
\begin{equation}
\int_{0}^{\infty}\mathrm{d}p f_{e}\left(p\right)=1.
\label{eq:npnorm}
\end{equation}
In the case of dark matter annihilation, $\beta_{e}=1.0$ to a very good 
approximation. Furthermore, $\gamma_{e}=E_{e}/m_{e}c^{2}$, which will be in the
range $19.6M_{\chi}<\gamma_{e}<1960M_{\chi}$, corresponding to energies
in the range $0.01-1.0\ M_{\chi}/\mathrm{GeV}$.
%
%
Eq~(\ref{eq:npspec}) can be written explicitly by retaining the power
law terms in Eq.~(\ref{eq:tylerne}) and their associated `weights'
$d_{i}$, and introducing a normalising factor $A=1/172.77$. The
momentum spectrum is then written
%
%
\begin{equation}
f_{e}\left(p\right)=A\sum_{i=1,6}d_{i}\left(\frac{m_{e}}{M_{\chi}}\right)^{1-\alpha_{i}}p^{-\alpha_{i}},
\label{eq:weightsum}
\end{equation}
where $\alpha_{i}=\left(2.5,1.5,1.0,0.5,-1.0,-2.0\right)$. Finally
then, we have
\begin{equation}
\frac{\mathrm{d}n_{e}}{\mathrm{d}p}=n_{e}\left(r\right)A\sum_{i=1,6}d_{i}\left(\frac{m_{e}}{M_{\chi}}\right)^{1-\alpha_{i}}p^{-\alpha_{i}}.
\label{eq:npnormspec}
\end{equation}

\section{Observables}
\label{sec:simobs}

We now calculate the magnitude of dSZ radio emission, and the flux of
up-scattered CMB photons, using the model parameters for Draco as
described in Section \ref{sec:dsph}. The computations use an
approximate method based on the work of \citet{ek}. This holds to
first order in the electron optical depth, neglects multiple
scatterings and is valid for a single electron population only.
Throughout, we employ units where $\langle\sigma V\rangle_{A}$ is
measured in $10^{-26}\mathrm{cm^{3} s^{-1}}$, and $M_{\chi}$ in
$\mathrm{GeV}$. Distances such as $r_{s}$ and $r_{c}$ are taken in
$\mathrm{kpc}$, and $n_{\chi,0}$ is in $\mathrm{cm^{-3}}$.

\subsection{The Sunyaev-Zeldovich Effect}
\label{subsec:SZeffect}

The SZ effect is caused by inverse Compton scattering of CMB photons
off energetic electrons. On average, the photons gain energy,
shifting their spectrum to higher frequencies, and causing a
distortion in the CMB radiation field. This is commonly characterised
by the Compton $y$-parameter, the line-of-sight integral of gas
pressure through the electron cloud:
\begin{equation}
y=\int \sigma_{T}\frac{P_{e}\left(r\right)}{m_{e}c^{2}}\mathrm{d}l.
\label{eq:yparam1}
\end{equation}
We may also compute the integrated $y$-parameter, $Y$, over the solid
angle $\Omega$ of the dSph:
\begin{equation}
Y=\frac{\sigma_{T}}{m_{e}c^{2}}\int\int P_{e}\mathrm{d}l\mathrm{d}\Omega.
\label{eq:Yparam1}
\end{equation}
An integral over solid angle can be written as
$\mathrm{d}\Omega=\mathrm{d}A/r^{2}_{h}$, hence we can equivalently write
\begin{equation}
Y=\frac{\sigma_{T}}{m_{e}c^{2}r^{2}_{h}}\int P_{e}\mathrm{d}V,
\label{eq:Yparam2}
\end{equation}
where $r_{h}$ is the heliocentric distance to the dSph (assumed to
 be $80\mathrm{kpc}$ for Draco.
For the ultra-relativistic gas considered here, where
$E_{e}\sim\mathrm{GeV}$, we apply the relationship between pressure
and energy density $P_{e}=\epsilon/3$. The (dominant) kinetic energy density
 $\epsilon$ is obtained from \citep{ek}
\begin{equation}
\epsilon=n_{e}\left(r\right)\int^{M_{\chi}/m_{e}}_{0.01M_{\chi}/m_{e}}\mathrm{d}pf_{e}\left(p\right)\left(\sqrt{1+p^{2}}-1\right)m_{e}c^{2}.
\label{eq:epsilon}
\end{equation}
where again the upper limit is not too important, as the source function
falls rapidly to zero as the electron energy approaches the neutralino
rest mass energy. We already showed that $p>19.5695M_{\chi}$, and so even for a
low neutralino mass of $10\mathrm{GeV}$, $p^{2}\gg1$. In this case, we can 
simplify Eq.~(\ref{eq:epsilon}) to
\begin{equation}
\epsilon=n_{e}\left(r\right) m_{e}c^{2}\int^{M_{\chi}/m_{e}}_{0.01M_{\chi}/m_{e}}\mathrm{d}pf_{e}\left(p\right)p.
\label{eq:epsilon2}
\end{equation}
The pressure is then evaluated using Eqs~(\ref{eq:ner}), (\ref{eq:weightsum})
 and (\ref{eq:epsilon}), leading to:
\begin{equation}
P_{e}=1.585\times10^{-12}n^{2}_{\chi}\sigV \mathrm{\ J\ m^{-3}},
\label{eq:pbar2}
\end{equation}
Note that the neutralino mass is present only in the number density
term here.

We may now calculate the integrated $Y$ parameter, using this result
and Eq.~(\ref{eq:Yparam2}), for each dSph model in Section
\ref{sec:dsph}:
\beq
Y=3.974\times10^{-8}r^{3}_{s}r^{-2}_{h}n^{2}_{\chi,0}
\sigV\int^{r_{t}/r_{s}}_{0}\hat{r}^2 a^{2}\left(\hat{r}\right)\mathrm{d}\hat{r},
\label{eq:intY}
\eeq
where $r_{s}$ and $a\left(\hat{r}\right)$ should be replaced with $r_{c}$
and $b\left(\hat{r}\right)$ for the CPL models.

To derive temperature shifts, we will work in terms of the mean Compton 
parameter averaged over the dSph, i.e.
\beq
\bar{y}=\frac{Y}{\Omega},
\label{ybar}
\eeq
where $\Omega$ is the angular extent over which the $Y$ parameter is averaged 
in Eq.~({\ref{eq:Yparam1}). Once converted to temperature 
units, $\bar{y}$ measures the temperature decrement inside a telescope beam of 
angular size $\Omega$. We choose $\Omega$ to take three values, first to 
match the angular size of the whole dSph, secondly within a 1' beam, and 
finally within a 1'' beam.

For the assumed neutralino mass, annihilation electrons
are always ultra-relativistic. Since we expect these particles to
have energies $\sim \mathrm{GeV}$, the effect of such a non-thermal
electron population is to completely remove photons from the spectral
range of the CMB. The problem is one of electron number density: a low
neutralino mass increases the number density of electrons as
$M^{2}_{\chi}$, which raises the scattering probability
accordingly. It is clear that the relativistic SZ signature really
measures the electron number in the dSph. Since there is a one-to-one
correspondence between the number of electrons and the number of
annihilating neutralinos, $Y$ is a direct measure of the 
dSph mass.
\begin{table*} 
\begin{center}
\caption{List of $\bar{y}$ parameters, fractional intensity
shifts $\delta i$, and temperature decrements $\Delta T$ for both sets
of dSph models, assuming $M_{\chi}=10\ \mathrm{GeV}$ and
$\sigV=10^{-26}\ \mathrm{cm^{3}\ s^{-1}}$. All quantities are averaged over 
the angular extent of the dSph. Bracketed values correspond
to the different Milky Way models as described in Table
\ref{tab:chmtable1}. A heliocentric distance to Draco of
$80\mathrm{kpc}$ is assumed. For the intensity and temperature shifts,
we use a frequency of $x=0.616$ i.e. 35GHz.
\label{tab:voltable}}
\begin{tabular}{cccc}
\hline \hline%
$\gamma$ & $\bar{y}/10^{-11}$ & $\delta i/10^{-13}$ & $\Delta T/10^{-13}$K\\
\hline
0.5 & 7.019 (1.338$\times10^{2}$) & 8.026 (1.530$\times10^{2}$) & -5.973 (-1.103$\times10^{2}$)\\
1.0 & 7.095 (1.336$\times10^{2}$) & 8.114 (1.528$\times10^{2}$) & -6.038 (-1.137$\times10^{2}$) \\
1.5 & 4.514$\times10^{1}$ (9.365$\times10^{2}$) & 5.162$\times10^{1}$ (1.071$\times10^{3}$) & -3.841$\times10^{1}$ (-7.970$\times10^{2}$) \\
\hline \hline
$\alpha$ & $\bar{y}/10^{-11}$ & $\delta i/10^{-13}$ & $\Delta T/10^{-13}$/K\\
\hline
0.2 & 4.045 (9.040$\times10^{1}$) & 4.626 (1.034$\times10^{2}$) & -3.442 (-7.693$\times10^{1}$) \\
0.0 & 2.970 (8.599$\times10^{1}$) & 3.396 (9.833$\times10^{1}$) & -2.527 (-7.318$\times10^{1}$) \\
-0.2 & 2.250 (7.586$\times10^{1}$) & 2.573 (8.676$\times10^{1}$) & -1.915 (-6.456$\times10^{1}$) \\
\hline \hline
\end{tabular}
\end{center}
\end{table*}

\subsubsection{The Intensity Shift}
\label{subsubsec:intensshift}

The fractional change in the CMB intensity field is
\begin{equation}
\delta i\left(x\right)=\frac{\Delta I\left(x\right)}{I_{0}},
\label{eq:intshift1}
\end{equation}
where $x=h\nu/k_{b}T_{\rm CMB}$ is the dimensionless frequency, and
$I_{0}=2\left(kT_{\rm
CMB}\right)^{3}/\left(hc\right)^{2}$. Contributions to such a
distortion are written as a product of a spectral function,
$g\left(x\right)$, and the Compton parameter $y$. The product
$g\left(x\right)y$ usually refers to the thermal electron population
in clusters of galaxies; for the relativistic gas considered here, we
write $\tilde{g}\left(x\right)$ to make the distinction
explicit. The fractional distortion averaged over the whole dSph is therefore
written as~\citep{raph}
\begin{equation}
\delta i\left(x\right)=\tilde{g}\left(x\right)\bar{y}.
\label{eq:intshift2}
\end{equation} 
There are two contributions to $\tilde{g}\left(x\right)$. First,
photons are removed from the infinitesimal frequency band $x,x+\Delta
x$ by collisions with the ultra-relativistic electrons. This
contribution is written $-i\left(x\right)$. The second effect is the
photons scattered into this band from lower frequencies, which is
written $j\left(x\right)$. Scattered CMB photons have their frequency
increased (on average) by a factor of $4\gamma^{2}_{e}/3-1/3$, so CMB
photons are up-scattered to the X-ray regime. Therefore, we are
interested in two distinct frequency bands - that near $x\sim2.5$ in
the radio corresponding to $i\left(x\right)$, and the band close to
$x\sim10^{6}$ in X-rays, described by $j\left(x\right)$.

Essentially no photons are scattered into the radio frequency band
under consideration from lower frequencies. Photons are simply removed
from the spectrum, causing a decrease in the number of photons, and
thus a corresponding decrease in the specific intensity. The factor
$i\left(x\right)$ therefore has the spectral form of the CMB
$i\left(x\right)=x^{3}/\left(e^{x}-1\right)$, which is maximal at
$x=2.82$.

The spectral factor $\tilde{g}\left(x\right)$ is thus given by~\citep{ek}
\begin{equation}
\tilde{g}\left(x\right)=\bigl[j\left(x\right)-i\left(x\right)\bigr]\frac{m_{e}c^{2}}{k_{b}\tilde{T_{e}}},
\label{eq:gtilde}
\end{equation}
We define the
pseudo-temperature $k_{b}\tilde{T_{e}}$, as the ratio of the gas
pressure $P_{e}$ to the electron number density $n_{e}$. For a thermal
electron distribution, this is equal to the thermodynamic temperature. In this 
case, we have
\begin{equation}
k_{b}\tilde{T_{e}}\equiv \frac{P_{e}}{n_{e}}=0.0122426\left(\frac{M_{\chi}}{\mathrm{GeV}}\right)\ \mathrm{GeV}.
\label{eq:kTeq}
\end{equation}
Using Eqns~(\ref{eq:intshift2}) and (\ref{eq:gtilde}), the fractional
distortion in the CMB at radio wavelengths, integrated over the dSph,
is therefore:
\beq
\delta i\left(x\right)=-i\left(x\right)\frac{m_{e}c^{2}}
{k_{b}\tilde{T_{e}}}\bar{y}.
\label{eq:idisttot1}
\eeq
The approximation of dropping $j(x)$ is valied provided $x
<10$~\citep{ek}.

\subsubsection{The Temperature Shift}
\label{subsubsec:tempshift}
The above result can equally be expressed as a temperature
shift. Expressing the result in this manner has the advantage of a
direct comparison to typical CMB telescope noise temperatures.
\beq \Delta T=\Bigl|\frac{\partial
I\left(x\right)}{\partial T}\Bigr|^{-1}\Delta I
\label{eq:deltaT1}
\eeq
gives the expected temperature shift in the CMB, where
$I\left(x\right)=I_{0}i\left(x\right)$. The partial derivative is
\beq
\frac{\partial I\left(x\right)}{\partial T}=I_{0}\frac{x^{4}e^{x}}{\left(e^{x}-1\right)^{2}},
\label{eq:bbderiv}
\eeq
and in conjunction with Eqs.~(\ref{eq:intshift1}),
(\ref{eq:idisttot1}) and (\ref{eq:deltaT1}), we have
\beq
\Delta T\left(x\right)=-\frac{e^{x}-1}{xe^{x}}\frac{m_{e}c^{2}}{k_{b}\tilde{T_{e}}}\bar{y}T_{CMB}.
\label{eq:deltaT2}
\eeq
Table \ref{tab:voltable} shows the results for the fiducial case of
$M_{\chi}=10\ \mathrm{GeV}$ and $\sigV = 10^{-26}\ \mathrm{cm^{3}\
s^{-1}}$, assuming a heliocentric distance $r_{h}=80\ \mathrm{kpc}$
for Draco and with a beam size that matches the angular size of the
dSph. We assume an observing frequency of $35$GHz ($x=0.616$, close to
Band 1 of the forthcoming Atacama Large Millimetre Array
(ALMA)). While such $\bar{y}$ values are prohibitively low for current
or near-future experiments -- for example, the upcoming South Pole
Telescope~\citep{ruhl} will only achieve noise temperatures of $\sim
\mu\mathrm{K}$ on arcminute scales. However, noise temperatures of
order $\sim10^{-9}\ \mathrm{K}$ may well be reached by a future
generation of radio/sub-mm telescopes, such as ALMA. Finally, Tables
\ref{tab:voltable1}-\ref{tab:voltable3} show the same quantities, but
for beam sizes of $1'$, $1''$ and $0.1''$, encompassing a range of
target specifications for ALMA.

It is instructive to compare the results presented here with
those of Colafrancesco (2004) on clusters of galaxies. Although
clusters are considerably larger objects, their central dark matter
number density is at least a factor of 10 lower than in the dSph
models considered here. Since the electron density is proportional to
$n^{2}_{\chi}$, we therefore have a relative increase in dSZ pressure
of at least a factor of 100 in dSphs. Using Eq~(\ref{eq:yparam1}), it
is also clear that the integral along the line of sight will be
proportional to the scale length of the object, of order $10^{2}\
\mathrm{kpc}$ for a cluster and $10^{-1}\ \mathrm{kpc}$ for
dSphs. Hence, these two competing factors almost cancel, so the
surface brightness for the dSZ effect is roughly equal for clusters
and dSphs.

In the most optimistic cases, the dSZ effect is within the grasp
of ALMA. For example, using the $\gamma=1.5$ cusped model at $100$GHz
(i.e. ALMA Band 3) and converting to brightness temperatures $T_{B}$
rather than the thermodynamic temperatures quoted in the Tables, the
ALMA sensitivity calculator for 64 dishes with resolution $1''$ gives
a signal-to-noise of unity in two hours. We present theoretical
constraints in the $M_{\chi}-T_{B}$ plane in
Fig.~\ref{fig:paramspace}. The plot displays the brightness
temperature for cusped models observed in a $1''$ beam with 24hr and 1
yr integration times at $100 \mathrm{GHz}$.
This plane contains a substantial portion of the parameter space
considered viable in particle physics. It is only for the most extreme
cusp ($\gamma=1.5$) and low neutralino masses that a detectable signal
is predicted.

\begin{figure}
\begin{center}
\includegraphics[width=7.0cm,angle=270]{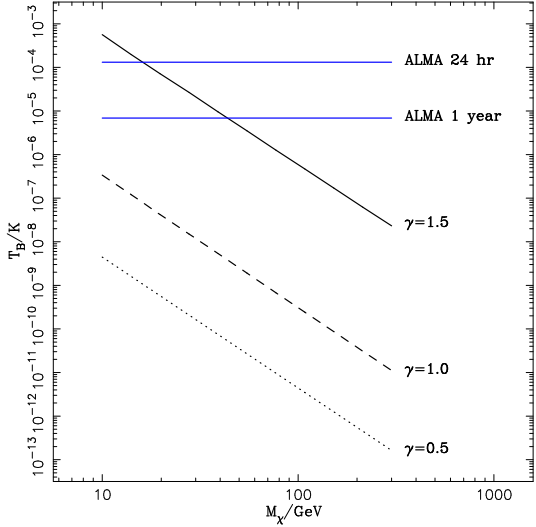} 
\caption{$M_{\chi}-T_{B}$ plane for three cusped dSph profiles, with an assumed 
NFW profile for the Milky Way (isothermal power-law models give almost 
identical results). We assume a 1'' beam at 100GHz, and display the brightness
temperature sensitivity for 24 hr and 1 yr ALMA observation.}
\label{fig:paramspace}
\end{center}
\end{figure}

However, we caution that the situation examined here may be
over-simplified for the cusped models. If such density profiles really
are followed to small radii, the Coulomb term neglected in
Eq. \ref{eq:bEhigh} may become important. This effect requires a more
thorough treatment to assess its impact -- although extra cooling
would lower the average electron energy, it would also drive the
electrons closer to a thermal equilibrium. The spectral factor would
in turn be closer to the thermal form, which tends to have larger
values at the frequencies considered here. The CMB temperature shifts
presented here may therefore be an underestimate at the cusp centre.

In addition, it is possible that the electron distribution may
deviate from the dark matter cusp structure once the Coulomb term
becomes dominant. In that case, the electron density distribution may
become less steep, and more like a cored profile. On the one hand,
this means that the peak signal will be lower. On the other, the
signal averaged over a larger beam increases, as the effective cusp
more closely matches the beam size. The tabulated temperature
decrements increase with decreasing resolution precisely for this
reason, and so a `smoother' cusp is a better match to the angular
scales of the telescopes mentioned above.

\begin{table*} 
\begin{center}
\caption{List of $\bar{y}$ parameters for 1' beam, fractional
intensity shifts $\delta i$, and temperature decrements $\Delta T$ for
both sets of dSph models. The remaining dSph parameters are as recorded in
Table~\ref{tab:voltable}. We assume $M_{\chi}=10\mathrm{GeV}$ and 
$\sigV=10^{-26}\mathrm{cm^{3}s^{-1}}$. For the intensity and temperature 
shifts, we use a frequency of $x=0.616$ i.e. 35GHz.\label{tab:voltable1}}
\begin{tabular}{cccc}
\hline \hline
$\gamma$ & $\bar{y}/10^{-7}$ & $\delta i/10^{-10}$ & $\Delta T/10^{-9}$K\\
\hline
0.5 & 3.209 (3.208) & 3.670 (3.590) & -2.731 (-2.671)\\
1.0 & 9.619 (9.582) & 1.100$\times10^{1}$ (1.096$\times10^{1}$) & -8.186 (-8.155)\\
1.5 & 5.119$\times10^{2}$ (5.499$\times10^{2}$) & -5.854$\times10^{2}$ (6.289$\times10^{2}$) & -4.356$\times10^{2}$ (-4.680$\times10^{2}$) \\
\hline \hline
$\alpha$ & $\bar{y}/10^{-8}$ & $\delta i/10^{-10}$ & $\Delta T/10^{-10}$/K\\
\hline
0.2 & 4.399 (4.397) & 5.031 (5.029) & -3.744 (-3.742) \\
0.0 & 4.473 (4.470) & 5.115 (5.112) & -3.806 (-3.804) \\
-0.2 & 4.418 (4.414) & 5.053 (5.048) & -3.760 (-3.757) \\
\hline \hline
\end{tabular}
\end{center}
\end{table*}

\begin{table*} 
\begin{center}
\caption{The same as Table~\ref{tab:voltable1}, but for a 1'' beam.
\label{tab:voltable2}}
\begin{tabular}{cccc}
\hline \hline
$\gamma$ & $\bar{y}/10^{-6}$ & $\delta i/10^{-9}$ & $\Delta T/10^{-9}$K\\
\hline
0.5 & 1.066 (1.057) & 1.219 (1.209) & -9.075 (-8.999) \\
1.0 & 8.102$\times10^{1}$ (7.251$\times10^{1}$) & 9.266$\times10^{1}$ (8.292$\times10^{1}$) & -6.895$\times10^{2}$ (-6.171$\times10^{2}$) \\
1.5 & 1.356$\times10^{5}$ (1.466$\times10^{5}$) & 1.551$\times10^{5}$ (1.677$\times10^{5}$) & -1.154$\times10^{6}$ (-1.248$\times10^{6}$) \\
\hline \hline
$\alpha$ & $\bar{y}/10^{-8}$ & $\delta i/10^{-11}$ & $\Delta T/10^{-10}$/K\\
\hline
0.2 & 4.528 (4.471) & 5.178 (5.113) & -3.853 (-3.805) \\
0.0 & 4.624 (4.548) & 5.288 (5.201) & -3.935 (-3.871) \\
-0.2 & 4.599 (4.497) & 5.259 (5.143) & -3.914 (-3.827) \\
\hline \hline
\end{tabular}
\end{center}
\end{table*}

\begin{table*} 
\begin{center}
\caption{The same as Table~\ref{tab:voltable}, but for a 0.1'' beam.
\label{tab:voltable3}}
\begin{tabular}{cccc}
\hline \hline
$\gamma$ & $\bar{y}/10^{-6}$ & $\delta i/10^{-9}$ & $\Delta T/10^{-8}$K\\
\hline
0.5 & 1.636 (1.515) & 1.871 (1.732) & -1.392 (-1.289) \\
1.0 & 1.544$\times10^{3}$ (9.999$\times10^{2}$) & 1.766$\times10^{3}$ (1.143$\times10^{3}$) & -1.314$\times10^{3}$ (-8.505$\times10^{2}$) \\
1.5 & 7.893$\times10^{6}$ (9.879$\times10^{6}$) & 9.026$\times10^{6}$ (1.130$\times10^{7}$) & -6.717$\times10^{6}$ (-8.407$\times10^{6}$) \\
\hline \hline
$\alpha$ & $\bar{y}/10^{-8}$ & $\delta i/10^{-10}$ & $\Delta T/10^{-10}$/K\\
\hline
0.2 & 5.200 (4.605) & 5.946 (5.266) & -4.249 (-3.919) \\
0.0 & 5.440 (4.694) & 6.221 (5.368) & -4.630 (-3.995) \\
-0.2 & 5.601 (4.665) & 6.405 (5.335) & -4.767 (-3.970) \\
\hline \hline
\end{tabular}
\end{center}
\end{table*}

\subsection{Upscattered Photons}

\label{subsec:upscat}
We now consider explicitly the spectrum and flux of upscattered CMB photons, 
adopting the first-order, approximate approach described in \cite{ek}.

\subsubsection{Frequency Spectrum}
\label{subsubsection:energyspec}

The scattered spectrum of photons, $j\left(x\right)$, can be expressed as
\beq
j\left(x\right)=\int_{0}^{\infty}\mathrm{d}t P\left(t\right)i\left(x/t\right),
\label{eq:scatspec}
\eeq
where the photon redistribution function $P\left(t\right)$ gives the
probability of a photon being scattered to a frequency $t$ times
greater than its original frequency. If the electron momentum spectrum
is $f_{e}\left(p\right)\mathrm{d}p$, the photon redistribution
function is 
\beq P\left(t\right)=\int_{0}^{\infty}\mathrm{d}p
f_{e}\left(p\right)P\left(t;p\right).
\label{eq:redist1}
\eeq
In this expression, $P\left(t;p\right)$ is the redistribution function
for a monoenergetic electron distribution. This has an analytic form
\begin{align}
P\left(t;p\right)=&-\frac{3\vert1-t\vert}{32p^{6}t}\bigl[1+\left(10+8p^{2}+4p^{4}\right)t+t^{2}\bigr]\nonumber\\
&+\frac{3\left(1+t\right)}{8p^{5}}\Bigl[\frac{3+3p^{2}+p^{4}}{\sqrt{1+p^{2}}}\nonumber\\
&-\frac{3+2p^{2}}{2p}\left(2\mathrm{arcsinh}\left(p\right)-\vert\ln\left(t\right)\vert\right)\Bigr],
\label{eq:ptp}
\end{align}
with the condition that $P\left(t;\alpha,p_{1},p_{2}\right)=0$ if
$\vert\ln\left(t\right)\vert>2\mathrm{arcsinh}\left(p_{2}\right)$.  We
may apply this formalism to the normalised electron momentum spectrum
in Eq.~(\ref{eq:weightsum}). For such high electron energies, it is
more convenient to express Eq~(\ref{eq:ptp}) in terms of the
logarithmic frequency shift $s=\mathrm{ln}\left(t\right)$, in which
case:
\begin{equation}
P\left(s;p\right)=P\left(e^{s};p\right)e^{s}\mathrm{d}s.
\label{eq:tsconvert}
\end{equation}
$P\left(t\right)$ is plotted in Figure \ref{fig:redist}, for
$M_{\chi}=10\mathrm{GeV}$ and $\sigV=10^{-26}\mathrm{cm^{3}\ s^{-1}}$
as before. Having computed this function, we calculate
$j\left(x\right)$ via Eq.~(\ref{eq:scatspec}). In Fig.\ref{fig:spec},
we plot $\vert j\left(x\right)-i\left(x\right)\vert$, which clearly
displays the prominent emission in the radio and X-ray bands. The two
regions are distinctly separated, with $i\left(x\right)$ constrained
to radio frequencies, while $j\left(x\right)$ dominating in
X-rays. There is virtually no overlap between the two. In addition,
$j\left(x\right)$ peaks close to $x\sim 2.0\times10^{5}$,
corresponding to $p\sim500$, as might be expected on the basis of the
power-law nature of Eqn.~(\ref{eq:weightsum}), and the expected
average electron momentum $p=4\gamma_{e}^{2}/3-1/3$.

\subsubsection{X-ray Flux}
\label{subsubsec:xrayflux}

The X-ray intensity produced by the upscattered photons can be
calculated from Eq.~(\ref{eq:intshift2}). In this instance, we replace
$\tilde{g}\left(x\right)\rightarrow j\left(x\right)$, and then the
X-ray flux density is 
\beq
F_{\rm Xray}\left(x\right)=I_{0}j\left(x\right)Y
\label{eq:xrayint1}
\eeq
where we integrate over the solid angle of the whole dSph.

We consider the X-ray emission integrated over a uniform efficiency
energy band $0.1-10\mathrm{keV}$ (corresponding to
$x_{1}=4.247\times10^{5}$ and $x_{2}=4.247\times10^{7}$), as an
approximation to the current X-ray satellites Chandra and
XMM-Newton. Dividing Eq.~(\ref{eq:xrayint1}) by $x$ and integrating
over the bandpass yields the X-ray photon flux
\beq
F_{\rm TOT}\left(x_{1},x_{2}\right)=\frac{I_{0}\tilde{Y}}
{kT_{\rm CMB}}\int_{x_{1}}^{x_{2}}\frac{j\left(x\right)}{x}\mathrm{d}x.
\label{eq:xrayflux1}
\eeq
For $M_{\chi}=10\ \mathrm{GeV}$ and $\sigV=10^{-26}\ \mathrm{cm^{3}\ s^{-1}}$,
 the integral above evaluates to 
\beq
\int_{x_{1}}^{x_{2}}\frac{j\left(x\right)}{x}\mathrm{d}x=0.414.
\label{eq:jint}
\eeq
The results for each of the models considered previously are listed in Table
\ref{tab:fluxtable}. The X-ray fluxes are all of order 
$10^{-12}\mathrm{cm^{-2}s^{-1}}$, above current estimates of the gamma-ray 
flux from direct annihilation channels, but well below what is currently 
possible to observe with X-ray satellites.

\begin{table} 
\begin{center}
\caption{List of integrated X-ray fluxes for Draco, in the energy band
$0.1-10\mathrm{keV}$, for $M_{\chi}=10\ \mathrm{GeV}$ and
$\sigV=10^{-26}\ \mathrm{cm^{3}\ s^{-1}}$.
\label{tab:fluxtable}}
\begin{tabular}{cc}
\hline \hline
$\gamma$ & $F_{\rm TOT}/10^{-12}\mathrm{cm^{-2}s^{-1}}$ \\
\hline
0.5 & 2.604 (2.564)\\
1.0 & 2.962 (2.914)\\
1.5 & 1.625$\times10^{1}$ (1.795$\times10^{1}$)\\
\hline \hline
$\alpha$ & $F_{\rm TOT}/10^{-12}\mathrm{cm^{-2}s^{-1}}$ \\
\hline
0.2 & 1.325 (1.301)\\
0.0 & 1.539 (1.436)\\
-0.2 & 1.955 (1.654)\\
\hline \hline
\end{tabular}
\end{center}
\end{table}

\begin{figure}
\begin{center}
\includegraphics[width=7.0cm,angle=270]{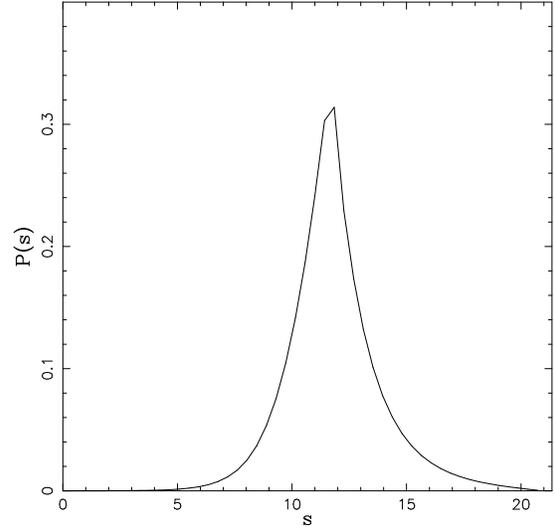} 
\caption{Photon redistribution function, for the momentum spectrum
Eq.~(\ref{eq:npnormspec}), assuming $M_{\chi}=10\ \mathrm{GeV}$.}
\label{fig:redist}
\end{center}
\end{figure}

\begin{figure}
\begin{center}
\includegraphics[width=7.0cm,angle=270]{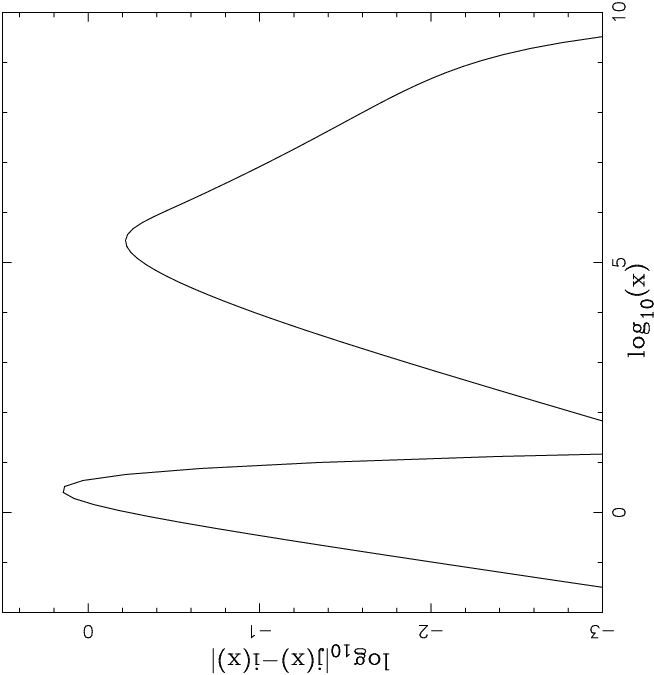} 
\caption{$\log_{10}\vert j\left(x\right)-i\left(x\right)\vert$ against
$\log_{10}\left(x\right)$, for the momentum spectrum
Eq.~(\ref{eq:npnormspec}). Note the two prominent bands, in the radio
near $x\sim2.5$ and in X-rays near $x=2.5\times10^{5}$.}
\label{fig:spec}
\end{center}
\end{figure}

\section{Conclusions}
\label{sec:conclusions}

In the above analysis, we have shown the following:

\begin{itemize}

\item{The Sunyaev-Zeldovich effect caused by secondary electrons
produced from dark matter annihilation in dwarf galaxies (the dSZ
effect) proposed by \cite{sc2004} could be measurable. The
Comptonisation parameters averaged over the angular size of the dSph
are $\bar{y}\sim10^{-11}$ for low neutralino masses of
$10\mathrm{GeV}$ and $\sigV=10^{-26}\mathrm{cm^{3}s^{-1}}$. The
temperature decrement for an assumed beam size of $1''$ is of the order
of milli-Kelvin for extremely cusped dSph halo models
and a few tenths of a nano-Kelvin for cored models. This may
provide a definitive test between these competing hypotheses, if the
signal for cusped model is large enough to be detectable by future
radio telescopes.  This result holds before the noisy effects of
primordial CMB, radio point sources, and SZ from clusters, has been
taken into account. This, however, is only a concern for cored models,
in which the signal comes from the bulk of the dSph. Even then, dSphs
are clean and uncontaminated objects, devoid of magnetic fields and
gas, and mostly free from point sources such as supernova
remnants. For cusped models, most of the signal comes from the very
centre and so contaminating point sources are not a worry. }

\item{Our main aim here has been to demonstrate the
feasibility of measuring the dSZ effect. We caution that our
calculations make use of an approximate, single-scattering formalism
that holds good for low electron optical depth. We may therefore have
underestimated the size of the effect in the very innermost regions of
cusped dSph models. Further numerical treatments, for example using the
methods of Colafrancesco et al. (2003), in the vicinity of
dark matter spikes are desirable.}

\item{Upscattered CMB photons lie in the X-ray band, with the emission
peak near $x=2.5\times10^{5}$ for the neutralino mass considered
here. Their integrated fluxes are $\sim10^{-12}\ \mathrm{cm^{-2}\
s^{-1}}$, comparable in size to that from gamma-rays produced by
direct annihilation channels. However, even next generation X-ray
satellites such as Constellation-X, with collecting areas of $\sim
1000\mathrm{cm}^{-2}$ will struggle to detect such a signal.}

\item{Our assessment of the importance of the dSZ effect is quite
optimistic. If dark haloes are strongly cusped, then we
conclude that the dSZ effect may be measurable in the near-future
by telescopes like ALMA. However, if dark haloes are only weakly
cusped, or if the dark matter particles are heavy ($\gta 50$ GeV),
then even the most generous integration times with ALMA may not yield
a positive detection. Nonetheless, it is worth bearing in mind that
here are some circumstances in which a larger effect may be
produced. First, the recently-discovered very dark dSph Ursa Major
\citep{willman,jan05} may be the first representative of the missing
dark satellites predicted by numerical simulations \citep{ben}. In
this case, there may be undetected, very dark dSphs much closer to us
than Draco, which is beneficial as the flux received obviously varies
like the inverse square of distance. Second, our calculations apply
only to the case of the neutralino dark matter candidate. There are
other possibilities, including light (1-5 MeV) scalar dark matter
\citep{boehm,hooper} and Kaluza-Klein dark matter \citep{bhs}, whose
induced dSZ signals could well be of interest.}

\end{itemize}

\section*{Acknowledgements}

We thank C. Tyler, T. Ensslin and K. Grainge for helpful discussions
and the anonymous referee for very useful comments. TLC acknowledges
support from a PPARC studentship.

\end{document}